# Analysis of the Relation between Artificial Intelligence and the Internet from the Perspective of Brain Science

Feng Liu[a,b,]*, Yong Shi [a,b, c,d,]*, Peijia Li[a,b]

[a]Research Center on Fictitious Economy and Data Science, the Chinese Academy of Sciences, Beijing 100190, China
[b]The Key Laboratory of Big Data Mining and Knowledge Management Chinese Academy of Sciences, Beijing 100190, China
[c]College of Information Science and Technology University of Nebraska at Omaha, Omaha, NE 68182, USA
[d]School of Economics and Management, University of Chinese Academy of Sciences, Beijing 100190, China

**Abstract**

Artificial intelligence (AI) like deep learning, cloud AI computation has been advancing at a rapid pace since 2014. There is no doubt that the prosperity of AI is inseparable with the development of the Internet. However, there has been little attention to the link between AI and the internet. This paper explores them with brain insights mainly from four views:1) How is the general relation between artificial intelligence and Internet of Things, cloud computing, big data and Industrial Internet from the perspective of brain science. 2) Construction of a new AI system model with the Internet and brain science.

## 1. How is the general relation between artificial intelligence and Internet of Things, cloud computing, big data and Industrial Internet from the perspective of brain science

A mass of new applications and features of the Internet have emerged in the past 20 years. For example, a printer or copying machine is remotely controlled; doctors perform operations through the remote network; Chinese water conservancy authorities place sensors in the soil, rivers and air so that the temperature, humidity, wind speed could be transmitted to the information processing center timely through the Internet, thus a report is formed, providing reference for decision-making in flood and drought control; Google launched the "Street View" service, with which, multi-lens cameras may be installed in a city so that the Internet users can enjoy the real-time scenes in Denver, Las Vegas, Miami, New York and San Francisco and other cities. It is not difficult to find that the structure of Internet is getting more and more similar to the brain structure.

These new Internet phenomena can be taken as the buddings of the motor nervous system, the somatosensory nervous system and the visual nervous system respectively. Based on the above new phenomena of the Internet, we published an article entitled *Findings and Analysis on the Law of the Internet Evolution* in September 2008, which analyzed the mature structure of the Internet from the perspective of neurology and abstracted it as an organizational structure that is highly similar to the human brain, namely the Internet virtual brain. It mainly focuses on finding and locating the position of the Internet's virtual hearing, visual, sensory, motor nervous systems and virtual central nervous system, etc. The brain-like structure chart of the Internet [1] is given in Fig.1.

* Corresponding author.
 *E-mail address:* zkyliufeng@126.com, yshi@ucas.ac.cn

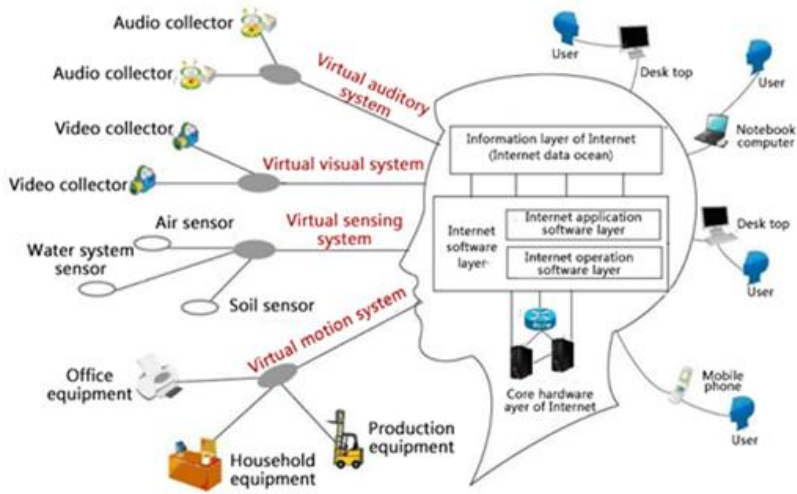

Fig. 1 Structure Diagram of the Internet Virtual Brain

From this figure, we could reanalyze the Internet of Things, cloud computing, big data, artificial intelligence (deep learning), Industry 4.0, Industrial Internet, unmanned aerial vehicles, intelligent driving and virtual reality, the following conclusions may be drawn.

*1.1. Internet of Things is the budding of the sensory nervous system of the Internet brain*

In November 2005, the International Telecommunication Union (ITU) released a report entitled *ITU Internet Reports 2005: The Internet of Things*, which formally proposed the item "Internet of Things (IOT)". The report did not define the Internet of Things exactly. But from a functional point of view, the ITU believes, that "all objects in the world can actively exchange information through the Internet, to achieve any interconnection between objects, omnipresent networks and omnipresent computing anytime and anywhere"; From a technical point of view, the ITU also believes that "the Internet of Things involves radio frequency identification technology (RFID)[2], sensor technology, nanotechnology and intelligent technology, etc."

Internet of Things highlights the concept of sensor perception, and it has the functions of network line transmission, information storage & processing, and industrial application interface, etc. Meanwhile, it often shares server, network lines and application interface with the Internet, which may be considered as the budding of the sensory nervous system of the Internet brain.

*1.2. Cloud computing is the budding of the central nervous system of the Internet brain*

After IBM and Google announced their cooperation in the field of cloud computing in October 2007, cloud computing has quickly become a hot issue in the industrial and academic research[3]. The birth of cloud computing has its historical roots. With the development of the Internet, the demand of the Internet's data storage capacity of emerging applications is growing and the Internet business is expanding as well. Therefore, the hardware and software maintenance costs of the Internet enterprises continue to increase, posing a heavy burden to most of them. At the same time, the Internet super-large enterprises such as Google, IBM and Amazon have a lot of spare hardware and software resources that are not fully utilized. In this case, It becomes a necessary trend that the Internet reforms from constructing the hardware and software separately to sharing the centralized cloud computing.

In the Internet virtual brain architecture, the central nervous system of the Internet virtual brain is the Internet's core hardware layer, which binds to the information layer to provide support and services for the virtual neural systems of the Internet. From the definition point of view, the cloud computing has consistent characteristics with the central nervous system of the Internet virtual brain. In an ideal situation, the sensor of Internet of Things or the Internet user interacted with cloud computing through the network lines and computer terminals, provide data to the cloud computing and accept the services.

*1.3. Industry 4.0, Industrial Internet, unmanned aerial vehicles, intelligent driving and 3D are essentially the budding of the Internet motor nervous system*

The Internet's central nervous system, namely the software system in cloud computing, controls the production equipment of industrial enterprises, household appliances, office equipment. It makes the machinery equipment such as intelligence, 3D printing and wireless sensor become the tools for the Internet brain to reform the world. At the same time, these intelligent manufactures and intelligent equipment also return data to the Internet brain so that the Internet central nervous system performs decision-making depending on such data. In the entire process, the technologies and applications such as Industry 4.0, Industrial Internet, unmanned aerial vehicles, intelligent driving, 3D printing are just the products from the development and budding of the Internet's motor nervous system.

*1.4. Big data is the basis of Internet brain information*

With the rise of technologies such as blogs, social networks, cloud computing, Internet of Things, and Industrial Internet, the data and information on the Internet are growing and accumulating at an unprecedented rate. Internet users' interaction with each other, information from enterprises and governments, and the real-time information from Internet of Things sensors are generating enormous amounts of structured and unstructured data at all times. Such data is scattered throughout the Internet network system in a huge volume. The data contains very valuable information on the economy, science and technology, education and so on. This is namely the background for the rise of the Internet data[4].

*1.5. AI is the foundation of Internet intelligence, wisdom and awareness*

Artificial intelligence, as the hottest issue in the Internet field since 2014, has been widely concerned by the scientific community, business and media. It was initially proposed as a concept by a group of far-sighted young scientists headed by McCarthy, Minsky, Rochester and Shennong in the summer of 1956 when they got together to study and explore the simulation of intelligence using machine and a series of other related issues.

In fact, the development of artificial intelligence is full of ups and downs. In the past 60 years, artificial intelligence has experienced many stages from optimism to pessimism, the climax to the low ebb. The latest low ebb occurred in 1992 when Japan's fifth-generation computer program was fruitless. Thereafter, the hotness in artificial neural network was cooled down at the beginning of the 1990s and the artificial intelligence field entered the "AI Winter" once again. That winter was so cold and long till 2006 when professor Geoffrey Hinton from Canadian University of Canada proposed "deep learning" algorithm, the situation was changed.

This algorithm is an ingenious upgrade to the artificial neural network theory, which was born in the 1940s The greatest innovation is the ability to process enormous amounts of data efficiently. Luckily, the feature was combined with the Internet, which led to a new upsurge of artificial intelligence since 2010. In 2011, a NCAP researcher and Andrew Ng from Stanford, namely Wu Enda who became the chief scientist of Baidu brain later, established Google brain based on deep learning. In 2013, Geoffrey Hinton joined Google, aiming to conduct Google brain more deeply.

After that artificial intelligence entered a new era -- the era of Internet artificial intelligence. Based on the mass "big data" of the Internet and its information exchange with the real-world, Baidu brain, Hyper brain and other Internet artificial intelligence systems also have emerged since 2014, and then new areas and records are continuously created.

From the above research, whether the Internet of Things, cloud computing, big data, Industry 4.0, Industrial Internet, unmanned aerial vehicles, intelligent driving, virtual reality or artificial intelligence (deep learning), they are still the products in the development process of the Internet, and they should be studied and considered in the large scale of the Internet evolution[5] (Fig.2).

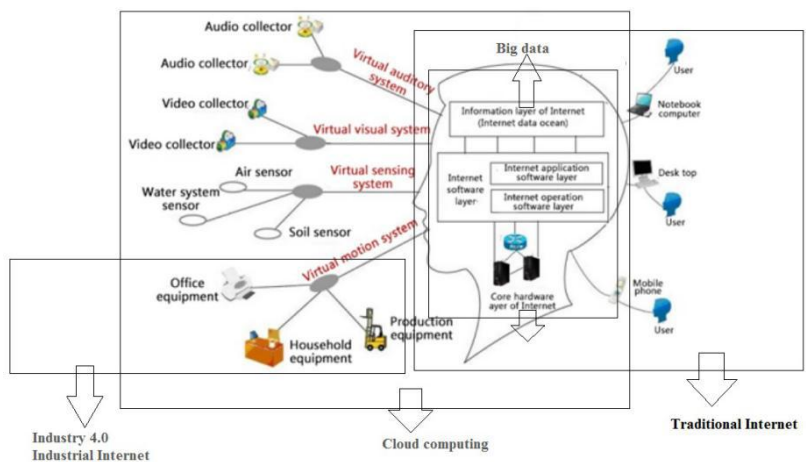

**Fig.2 the Internet of Things, cloud computing, big data, artificial intelligence in Internet's virtual brain**

## 2. Construction of a new AI system model with the Internet and brain science

*2.1. The brain-like Internet architecture provides new insights for the establishment of new artificial intelligence systems*

The Internet born in the last century has brought more and more influence on the human society. Various signs indicate that the Internet has a close relation with brain science. From 2005, the research suggests that "the Internet will evolve toward the state highly similar to the human brain. It will not only have its own visual, auditory, tactile,, motor nervous systems, but also have its own memory nervous system, central nervous system and autonomic nervous system. On the other hand, the human brain has evolved all the Internet functions at least tens of thousands of years ago, and the continuously-developing Internet will help neurologists to reveal the secret of the brain."

This research inspires us to explore if it is possible to build the artificial intelligence system model simulating the Internet brain in a supercomputer by the Internet function and architecture Meanwhile it need us to check if it is also possible to increase or reduce relevant functions and architectures in the building process according to the latest development of the Internet. The software system of this model can also be integrated in the chip.

*2.2. The implementation of the artificial intelligence system with the Internet-like brain*

The theoretical basis of the artificial intelligence system with the Internet-like brain is to implement the above-mentioned brain-like model of the Internet in the supercomputer. Besides it should combine the stable typical application and architecture of the Internet, simulate with the program and database, and present them in a visual form.

As a huge system, the Internet, after experiencing nearly 45 years of development, has already contained thousands of applications and subsystems. In addition, as the result of rapid development, the Internet has new applications every day. When using the Internet brain model, it is hard to contain all the Internet applications and subsystems in practice.

Therefore, we proposed to conduct the test by establishing the application library of the artificial intelligence system with the Internet-like brain (IBML) using the selected relatively-mature Internet applications with a high penetration rate. The applications in the application library of the artificial intelligence system model of the Internet-like brain may be added or reduced regularly according to the development of the Internet. This can avoid the problems like numerous Internet applications and frequent simultaneous disappearance and emergence. For example, we can establish an application library of the artificial intelligence system model of the Internet-like brain which is similar to the one below:

IBML= {Google, Facebook, Blogger, Amazon, Yahoo, YouTube, Quora, Wikipedia, Twitter, IPv4/IPv6...}

Through the establishment of the application library of the artificial intelligence system model of the Internet-like brain, it is possible to establish a new artificial intelligence system model by imitating the Internet function and architecture in the supercomputer. The establishment of the artificial intelligence system model of the Internet-like brain includes the following three steps:

**Step I. Hardware basis for the artificial intelligence system of the Internet-like brain**: 1) Large-size computer; 2) Laboratory-level sensor network. The sensor network will interact with the other parts of the brain model in the "neuron" social network account in the Internet brain-like model.

**Step II. Realization of the functions of the artificial intelligence system of the Internet-like brain**

1. Establishing the micro-social network, Wikipedia and search engine functions in this artificial intelligence system of the Internet-like brain

2. Connecting the virtual visual, auditory, sensory and motion systems built in the micro Internet of Things into the artificial intelligence system of the Internet-like brain

3. Operating the micro-social network, Wikipedia and search engine, as well as the micro Internet of Things system built in the artificial intelligence system model of the Internet-like brain to generate big data

4. Applying the artificial intelligence algorithms like machine learning and deep learning to the artificial intelligence system model of the Internet-like brain (Fig.3).

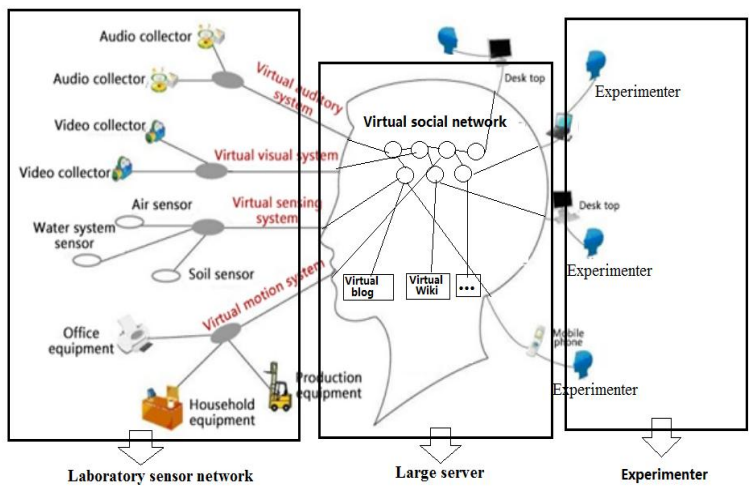

Fig.3 the artificial intelligence system model of the Internet-like brain

**Step III. Present the operation of the artificial intelligence system of the Internet-like brain in the server in a visual way by using data visualization technology as the interactive interface so that people can control it by neuron accounts simulating social network.**

Study method: Continue to improve, update and study the artificial intelligence system model of the Internet-like brain, and add new functions in the Internet and brain science to the model to conduct testing; continue to add information and knowledge base systems in the artificial intelligence system model of the Internet-like brain; perform the artificial intelligence processing and experiment personnel active operation processing on each neuron simulating social network; and observe the intelligent characteristics of the artificial intelligence system model of the Internet-like brain.

## 3. Conclusion:

The combination of Internet, brain science and artificial intelligence is related a number of fields and has some influence on them, including smart city, artificial intelligence, Internet of Things, cloud computing, big data, robots, Industrial Internet, brain science, philosophy of science and technology and etc. The new achievements in these fields will be introduced in our papers in the future.